\documentclass[twocolumn,letterpaper]{IEEEAerospaceCLS}  

\usepackage[]{graphicx}    
\usepackage[colorlinks=true, urlcolor=blue, linkcolor=black, citecolor=black]{hyperref} 
\newcommand{\ignore}[1]{}  
\usepackage{amssymb}
\usepackage{amsmath}
\usepackage{amsfonts}
\usepackage{amsbsy}
\usepackage{cite}
\usepackage[mathscr]{euscript}
\let\euscr\mathscr \let\mathscr\relax
\usepackage{fancyvrb}
\usepackage{fvextra}
\usepackage{hyphenat}

\newcommand{\eqdef}{\triangleq}
\newcommand{\bm}[1]{\mathbf{#1}}
\newcommand{\diag}{\text{diag}}
\newcommand{\Cset}{\mathbb{C}}

\newcommand{\Zset}{\mathbb{Z}}
\newcommand{\be}{\begin{equation}}
\newcommand{\ee}{\end{equation}}
\newcommand{\bb}{\mathbf{b}}

\newcommand{\trasp}{\text{T}}
\newcommand{\herm}{\text{H}}
\newcommand{\Es}{\mathbb{E}}

\newcommand{\Hb}{\mathbf{H}}

\usepackage{tikz}

\newcommand\copyrighttext{%
  \footnotesize \textcopyright \the\year{} IEEE. Personal use of this material is permitted. Permission from IEEE must be obtained for all other uses, including reprinting/republishing this material for advertising or promotional purposes, collecting new collected works for resale or redistribution to servers or lists, or reuse of any copyrighted component of this work in other works.}

\newcommand\acceptedtext{%
  \footnotesize This article has been accepted for publication in the proceedings of this conference, but has not been fully edited. \\Content may change prior to final publication. \\
}
  
\newcommand\acceptednotice{%
\begin{tikzpicture}[remember picture,overlay]
\node[anchor=north,yshift=0pt,xshift=0pt] at (current page.north) {%
\begin{minipage}{\textwidth}
\center \acceptedtext
\end{minipage}};
\end{tikzpicture}%
}

\begin{document}
\title{AI-Driven Design of Stacked Intelligent Metasurfaces for Software-Defined Radio Applications}

\author{%
Ivan Iudice\\ 
Security Unit\\
Italian Aerospace Research Centre (CIRA)\\
Capua I-81043\\
i.iudice@cira.it
\thanks{\footnotesize 979-8-3315-7360-7/26/$\$31.00$ \copyright2026 IEEE}              
}

\author{%
Ivan Iudice\\ 
Security Unit\\
Italian Aerospace Research Centre\\ 
(CIRA)\\
Capua I-81043\\
i.iudice@cira.it
\and
Giacinto Gelli\\
Dept. of Electrical Engineering\\
and Information Technology \\
University of Naples Federico II\\
Naples I-80125\\
gelli@unina.it
\and
Donatella Darsena\\ 
Dept. of Electrical Engineering \\
and Information Technology\\
University of Naples Federico II\\
Naples I-80125\\
darsena@unina.it
\thanks{\footnotesize 979-8-3315-7360-7/26/$\$31.00$ \copyright2026 IEEE.
Personal use is permitted, but republication/redistribution requires IEEE permission.
See https://www.ieee.org/publications/rights/index.html for more information}              
}

\maketitle
\acceptednotice

\thispagestyle{plain}
\pagestyle{plain}

\begin{abstract}
The integration of reconfigurable intelligent surfaces (RIS) into future wireless communication systems offers promising capabilities in dynamic environment shaping and spectrum efficiency. In this work, we present a consistent implementation of a stacked intelligent metasurface (SIM) model within the NVIDIA’s AI-native framework Sionna for 6G physical layer research. Our implementation allows simulation and learning-based optimization of SIM-assisted communication channels in fully differentiable and GPU-accelerated environments, enabling end-to-end training for cognitive and software-defined radio (SDR) applications. We describe the architecture of the SIM model, including its integration into the TensorFlow-based pipeline, and showcase its use in closed-loop learning scenarios involving adaptive beamforming and dynamic reconfiguration. Benchmarking results are provided for various deployment scenarios, highlighting the model’s effectiveness in enabling intelligent control and signal enhancement in non-terrestrial-network (NTN) propagation environments. This work demonstrates a scalable, modular approach for incorporating intelligent metasurfaces into modern AI-accelerated SDR systems and paves the way for future hardware-in-the-loop experiments.
\end{abstract}

\tableofcontents

\section{Introduction}

Sixth-generation (6G) wireless communication systems are expected to dramatically outperform previous generations by leveraging AI-native technologies, the \emph{non-terrestrial network} (NTN) paradigm, software-defined radio architectures, and programmable radio environments \cite{Let2019}. A particularly promising approach in this context is that of \emph{reconfigurable intelligent surfaces} (RIS), which enable the realization of \emph{software-defined environments} by dynamically shaping the propagation medium and enhancing spectrum and energy efficiency through programmable manipulation 
of incident electromagnetic (EM) waves \cite{Pan2020,Bas2021,Worka2024}.

Beyond the substantial progress made in RIS research, recent studies have investigated multi-layer and dynamic implementations, so called 
\emph{stacked intelligent metasurfaces} (SIMs), to further increase the available degrees-of-freedom (DoF). Inspired by diffractive optical platforms, SIMs are realized as cascades of programmable metasurfaces, each endowed with adaptive or intelligent control capabilities \cite{Hassan.2024, Nerini.2024, DiRenzo, Hanzo}. SIMs can implement signal-processing transformations directly in the EM wave domain, generating tailored wave profiles as the input propagates through the layers. Their integration into wireless systems can replace conventional digital beamforming architectures, reducing the need for high-resolution DACs/ADCs and minimizing the number of RF chains, thereby lowering hardware costs and power consumption. Moreover, since multi-antenna precoding and combining occur directly in the wave domain at the speed of light, SIM architectures mitigate processing delays of digital solutions.

To the best of our knowledge, despite these advances, existing studies rarely consider key aspects such as scalability, differentiability, and integration into AI-native physical-layer pipelines. Moreover, only limited attention has been given in the metasurface literature to programmable NTN scenarios or GPU-accelerated architectures suitable for closed-loop learning.
These gaps motivate the development of a framework aimed at enabling scalable simulation and optimization of multi-layer programmable surfaces in realistic NTN contexts.

To this end, we propose the first consistent integration of a SIM model within NVIDIA’s AI-native framework \emph{Sionna}. Sionna is a GPU-accelerated, fully differentiable library for communication system simulation and optimization, built on TensorFlow and designed for rapid prototyping and end-to-end learning in next-generation wireless systems \cite{Hoy2022,SionnaSW2025}. Our SIM implementation extends Sionna’s modular architecture, enabling efficient simulation and gradient-based optimization of SIM-assisted channels in both terrestrial and NTN environments.

The main contributions of this paper are summarized as follows:
\begin{itemize}
  \item We develop a mathematically grounded and differentiable model of SIM, designed for seamless integration into Sionna’s TensorFlow-based simulation pipeline.
  \item We demonstrate efficient GPU-accelerated simulation and optimization of SIM, exploiting Sionna’s end-to-end gradient propagation and modular design.
  \item We apply our framework to a closed-loop learning scenario, exemplified by adaptive beamforming.
\end{itemize}

We validate the proposed framework through numerical benchmarks, comparing SIM-enabled systems with conventional multi-antenna baselines. The obtained results highlight significant improvements in spectral efficiency and signal fidelity. 
Overall, this work establishes a scalable foundation for integrating intelligent metasurfaces into AI-driven SDR systems and paves the way towards hardware-in-the-loop experimentation and real-world deployment.

The remainder of the paper is organized as follows. Section~\ref{sec:background} reviews related work on RIS and differentiable simulation frameworks. Section~\ref{sec:system-model} introduces the mathematical model of stacked intelligent metasurfaces, while Section~\ref{sec:model-opt}  describes the model-based optimization. 
Section~\ref{sec:implementation} details the implementation within Sionna. Section~\ref{sec:experiments} presents learning-based optimization and performance benchmarks. Section~\ref{sec:conclusion} concludes the paper, discussing future research directions.

\section{Background and Related Work}
\label{sec:background}

Reconfigurable intelligent surfaces (RISs), also referred to as \emph{intelligent reflecting surfaces} (IRSs), have recently emerged as a fundamental paradigm for future 6G networks, providing the basis for programmable wireless environments \cite{Zha2019,Liu2020}.  
Composed of tunable sub-wavelength elements, RIS can manipulate incident EM waves by dynamically adjusting their reflection, refraction, or absorption characteristics. Compared to conventional active relays, they offer low-power operation and scalable deployment, and have been widely studied for many applications, 
such as energy-efficient communications, beamforming, secure transmission, and localization \cite{Pan2020,Put2025,Ume2023}. Moreover, RIS are receiving increasing attention in non-terrestrial networks (NTN), where satellite and aerial platforms can benefit from enhanced channel programmability \cite{Worka2024}.  

Beyond traditional single-layer designs, the concept of \emph{stacked intelligent metasurfaces} (SIMs) introduces multiple programmable layers with adaptive or intelligent capabilities \cite{Wu2021}. Such architectures enable richer channel transformations, finer wavefront control, and improved robustness against hardware constraints. By operating directly in the EM domain, SIMs can implement signal processing transformations in the wave itself, tailoring the input as it propagates through the layers to produce a desired output profile. This approach opens the way to replacing conventional digital beamforming structures, thereby reducing the need for high-resolution DACs/ADCs and minimizing the number of RF chains, with consequent savings in hardware cost and power consumption. Furthermore, multi-antenna precoding and combining are inherently performed in the wave domain as the signal propagates at the speed of light, reducing processing delay 
in comparison to purely digital solutions.  

Despite these promising features, the integration of SIMs into differentiable, AI-native simulation frameworks remains largely unexplored. In parallel, the rise of AI-driven physical-layer research has motivated the development of differentiable, GPU-accelerated simulators. Among them, NVIDIA’s \emph{Sionna} library has been introduced as a modular TensorFlow-based platform supporting automatic differentiation, end-to-end learning, and rapid prototyping \cite{Hoy2022}, while its recent extension Sionna RT incorporates differentiable ray tracing for realistic propagation modeling in dynamic environments \cite{NVlab2023}. Nevertheless, no scalable multi-layer SIM model has yet been integrated into Sionna, thus eaving an open gap for the simulation and optimization of intelligent metasurfaces in software-defined radio (SDR) and NTN scenarios.

\section{System Model}
\label{sec:system-model}

\begin{figure*}
    \centering
    \includegraphics[width=0.8\linewidth]{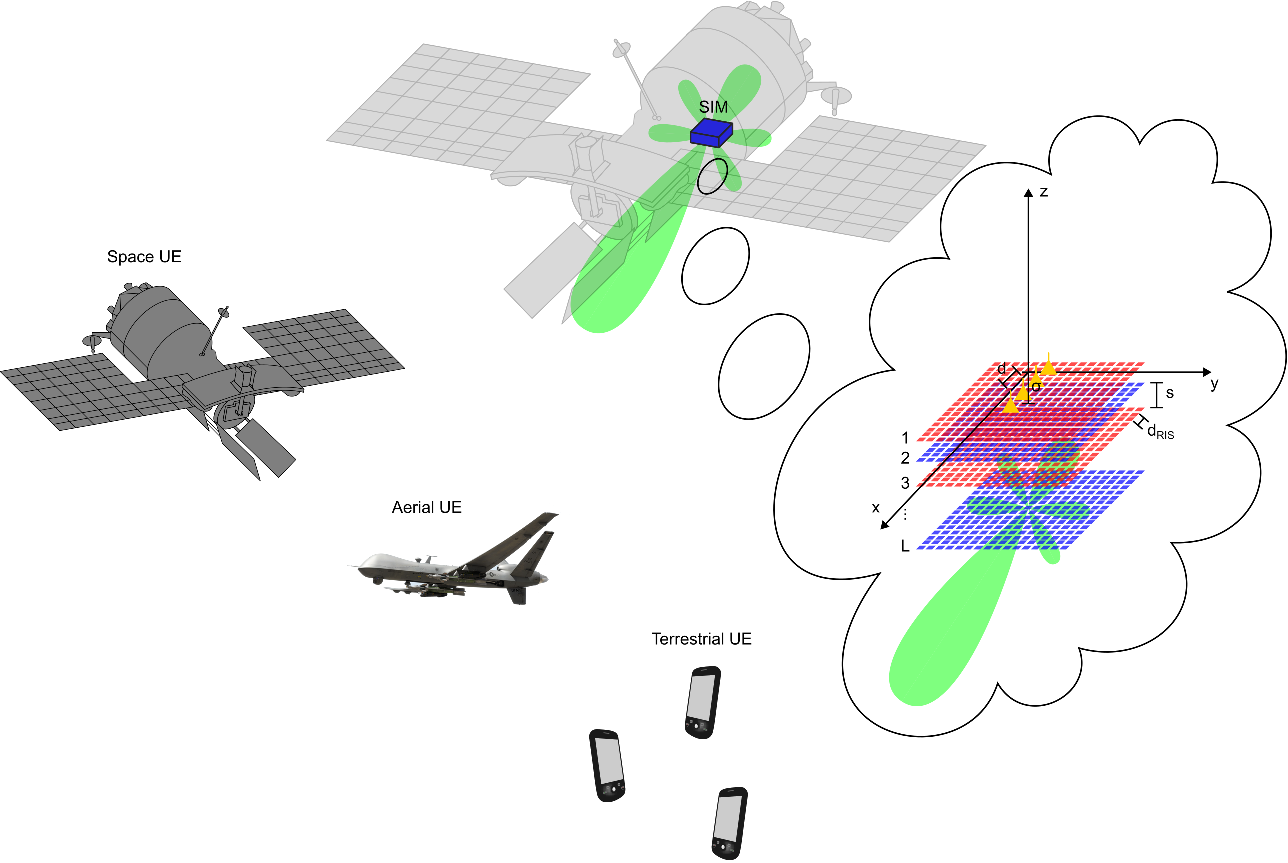}
    \caption{SIM-aided downlink architecture with amplitude-controlled (red) and phase-controlled (blue) layers.}
    \label{fig:fig_1}
\end{figure*}

We consider the SIM-aided downlink communication 
system illustrated in Fig.~\ref{fig:fig_1}.
In this setup, a satellite (SAT) is equipped with $N$ antennas
and employs a SIM consisting of $L$ planar metasurface layers to communicate with
$K$ single-antenna user equipments (UEs).
Let $s$ denote the spacing between two adjacent layers of the SIM,
and $\sigma$ the distance between the array and the first layer of the SIM.  
The $\ell$-th layer of the SIM is composed of
$Q^{(\ell)} \eqdef Q^{(\ell)}_{x} \times Q^{(\ell)}_{y}$
meta-atoms arranged in a rectangular grid, 
with $Q^{(\ell)}_{x}$ and $Q^{(\ell)}_{y}$
elements along the $x$ and $y$ axes, respectively,
and a constant inter-element spacing $d_{\text{RIS}}$.
For simplicity, we introduce a mapping that converts the 2-D index
$(q_x,q_y)$ of a meta-atom in the $\ell$-th layer,
where $q_x \in \left\{0, 1, \ldots, Q^{(\ell)}_{x}-1\right\}$
and $q_y \in \left\{0, 1, \ldots, Q^{(\ell)}_{y}-1\right\}$,
into a one-dimensional (1-D) index
$q \eqdef q_x \, Q^{(\ell)}_{y} + q_y$
belonging to $\mathcal{Q^{(\ell)}} \eqdef \{0, 1, \ldots, Q^{(\ell)}-1\}$. This mapping indexes the meta-atoms sequentially from row to row within each layer.
It should be noted that, within this architecture, each layer can be designed with a different number of meta-atoms, allowing flexibility in the overall metasurface configuration.

In this work, we adopt the well-established analytical propagation model proposed in \cite{Lin.2018,Liu.2022,Hassan.2024,Nerini.2024,DiRenzo,Hanzo,DiRenzo-ICC,Liu.2024,Yao.2024,Lin.2024}.
Specifically, each metasurface layer is assumed to be perfectly impedance-matched, thereby eliminating backward reflections and focusing exclusively on forward propagation.  
When an incident EM wave with carrier frequency $f_0>0$ impinges on a meta-atom of the first layer, the amplitude and phase of the transmitted wave are determined by the product of the incident electric field and the complex transmission coefficient of that meta-atom.  
The resulting transmitted wave acts as a secondary source that illuminates all meta-atoms of the subsequent layer, in accordance with the Huygens--Fresnel principle \cite{Goodman}.

Furthermore, to enhance the degrees-of-control (DoC) in the wave-domain transformation
performed by the SIM, we consider the model architecture proposed in \cite{Dar2025},
incorporating both \emph{amplitude-controlled} (AC) and \emph{phase-controlled} (PC) layers into the stacked device \cite{Dar2025}. 
Specifically, PC layers are nearly passive, allowing only phase adjustments of
their meta-atoms via components such as 
varactor or PIN diodes. In contrast, AC layers are active 
and enable amplitude modulation of their meta-atoms through the integration of amplifier chips. This combination  
of PC and AC layers in the SIM facilitates independent manipulation 
of amplitude and phase in the wave domain.

Let $\tau_{\ell,q} = \alpha_{\ell,q} \,  e^{j \phi_{\ell,q}}$ represent the EM transmission coefficient of the $q$-th meta-atom in the 
$\ell$-th metasurface layer, with $\ell \in \mathcal{L} \eqdef \{1, 2, \ldots, L\}$
and $q \in \mathcal{Q}^{(\ell)}$.
We define $\mathcal{L}_{\text{ac}}$ and 
$\mathcal{L}_{\text{pc}}$ as two nonoverlapping subsets of $\mathcal{L}$ 
that index the AC and PC layers, respectively, whose 
cardinalities $L_{\text{ac}}$ and $L_{\text{pc}}$ satisfy the condition 
$L_{\text{ac}} + L_{\text{pc}}=L$. 
The transmission coefficients for each layer of the SIM are organized into diagonal matrices 
$\bm T_{\ell} \eqdef \diag(\pmb \tau_\ell) \in \Cset^{Q^{(\ell)}
\times Q^{(\ell)}}$, where $\pmb \tau_\ell \eqdef \left[\tau_{\ell,0},
\tau_{\ell,1}, \ldots, \gamma_{\ell,Q^{(\ell)}-1}\right]$
and $\ell \in \mathcal{L}$.
AC layers consist of meta-atoms whose amplitude responses 
$\{\alpha_{\ell,q}\}_{\ell \in \mathcal{L}_{\text{ac}}}$ 
can be independently controlled through software.
The phases $\{\phi_{\ell,q}\}_{\ell \in \mathcal{L}_{\text{ac}}}$ 
of the transmission coefficients in AC layers are fixed and cannot be adjusted.
Consequently, these phases will be treated as known but uncontrollable
in the subsequent optimization process.
For PC layers, the metasurfaces are locally passive, i.e., 
their meta-atoms cannot amplify the incident EM waves. 
Due to the unavoidable material losses, PC layers 
may attenuate the EM waves that penetrate through them, implying that 
their amplitude responses are generally smaller than one, 
i.e., $\alpha_{\ell,q} \le 1$ for $\ell \in \mathcal{L}_{\text{pc}}$.
Therefore, we assume that the PC layers have a constant 
transmittance, meaning that the amplitude responses are fixed at
$\alpha_{\ell,q}=\alpha_\text{pc} \le 1$ 
for $\ell \in \mathcal{L}_{\text{pc}}$. 
The phases $\{\phi_{\ell,q}\}_{\ell \in \mathcal{L}_{\text{pc}}}$ can be adjusted within the interval $[0,2\pi)$.

The channel coefficients between the $N$ 
transmit antennas of the SAT array
and the $Q^{(1)}$ meta-atoms of the first layer of the SIM
are organized into the matrix $\bm W_1 \in \Cset^{N \times Q^{(1)}}$.
These coefficients are modeled using  {\em Rayleigh-Sommerfeld diffraction 
theory} as follows
\be
\{\bm W_1\}_{n,q} = \frac{A_{\text{bs}} \, \cos(\theta^{(1)}_{n,q})}{d^{(1)}_{n,q}}
\left(\frac{1}{2 \pi d^{(1)}_{n,q}}-\frac{j}{\lambda_0}\right) \, 
e^{j \frac{2 \pi}{\lambda_0} {d^{(1)}_{n,q}}}
\label{eq:W1}
\ee
for $n \in \mathcal{N} \eqdef \{0, 1, \ldots, N-1\}$
and $q \in \mathcal{Q}^{(1)}$, where $\lambda_0 = c/f_0$ is the wavelength, 
with $c = 3 \cdot 10^8$ m/s denoting the light speed in  vacuum, 
$A_{\text{bs}}$ is the effective area of the antennas of the array (evaluated at $f_0$), 
$\cos(\theta^{(1)}_{n,q})=\sigma/d^{(1)}_{n,q}$, and $d^{(1)}_{n,q}$ denotes the distance between the $n$-th antenna of the SAT and the $q$-th meta-atom of the first layer. 
This distance reads as
\be
d^{(1)}_{n,1} =  \sqrt{\left(x_q^{(1)}-x_n^{(0)}\right)^2 + \left(y_q^{(1)}-y_n^{(0)}\right)^2 +\sigma^2}
\label{eq:dqq-1}
\ee
with  $\left(x_q^{(1)}, y_q^{(1)}\right)$
and $\left(x_n^{(0)}, y_n^{(0)}\right)$ 
representing the $2$-D coordinates of the $q$-th meta-atom
on the first layer of the SIM and the position of the
$n$-th antenna of the SAT, respectively.

Similarly, for each $\ell \in \mathcal{L}-\{1\}$, 
the forward propagation process between layers $\ell-1$ and $\ell$ is 
described by the matrix 
$\bm W_\ell \in \mathbb{C}^{Q^{(\ell-1)} \times Q^{(\ell)}}$, whose 
entry $\{\bm W_\ell\}_{\tilde{q},q}$, for $\tilde{q} \in \mathcal{Q}^{(\ell-1)}$
and $q \in \mathcal{Q}^{(\ell)}$, represents
the channel coefficient from the $\tilde{q}$-th meta-atom
in the $(\ell-1)$-th layer to the $q$-th meta-atom
in the $\ell$-th layer, and is given by
\be
\{\bm W_\ell\}_{\tilde{q},q} = \frac{A_{\text{meta}} \,
\cos(\theta^{(\ell)}_{\tilde{q},q})}{d^{(\ell)}_{\tilde{q},q}} 
\left(\frac{1}{2 \pi d^{(\ell)}_{\tilde{q},q}}-\frac{j}{\lambda_0}\right) \, 
e^{j \frac{2 \pi}{\lambda_0} {d^{(\ell)}_{\tilde{q},q}}}
\label{eq:Well}
\ee
where $A_{\text{meta}}$ is the physical area of the meta-atoms, 
$\cos(\theta^{(\ell)}_{\tilde{q},q})={s}/{d^{(\ell)}_{\tilde{q},q}}$,
and $d^{(\ell)}_{\tilde{q},q}$ represents the propagation distance between 
the $\tilde{q}$-th meta-atom of the $(\ell-1)$-th layer
and the ${q}$-th meta-atom of the $\ell$-th layer. This distance is given by the following expression
\be
d^{(\ell)}_{\tilde{q},q} =  \sqrt{\left(x_q^{(\ell)}-x_{\tilde{q}}^{(\ell-1)}\right)^2 + 
\left(y_q^{(\ell)}-y_{\tilde{q}}^{(\ell-1)}\right)^2 + s^2} 
\label{eq:dqq}
\ee
with  $\left(x_q^{(\ell)}, y_q^{(\ell)}\right)$
representing the $2$-D coordinates of the $q$-th meta-atom 
on the $\ell$-th layer of the SIM.

The overall forward propagation through the SIM is described by the matrix
\be
\mathbf{G} = \mathbf{W}_1 \mathbf{T}_1 \mathbf{W}_2 \mathbf{T}_2 \cdots 
\mathbf{W}_{L-1} \mathbf{T}_{L-1} \mathbf{W}_L \mathbf{T}_L 
\in \mathbb{C}^{N \times Q^{(L)}}.
\label{eq:forward}
\ee


For the sake of simplicity, we assume that the SAT employs
for transmission a linear modulation format. In particular,
let $\{b_k[i]\}_{i \in \Zset}$ represent the data stream
to be transmitted to the $k$-th UE,
for $k \in \{0,1,\ldots,K-1\}$,
the complex envelope of the narrowband continuous-time 
signal associated with the $k$-th data stream is given by
\be
s_k(t) = \sum_{i=-\infty}^{+\infty} 
b_k(i) \, \psi(t-i \, T_\text s).
\label{eq:G}
\ee
where $\{b_k(i)\}_{k=0}^{K-1}$ are mutually independent sequences 
of zero-mean unit-variance independent and identically-distributed (i.i.d.) 
complex random variables.
These sequences are transmitted at symbol rate $1/T_\text s$, 
with $\psi(t)$ representing the real unit-energy square-root Nyquist 
pulse-shaping waveform.

Let the vector $\bm x(t) \eqdef [x_0(t), x_1(t),
\ldots, x_{N-1}(t)] \in \Cset^{1 \times N}$
denote the complex envelope of the signal transmitted
by the SAT antenna array, which can be written as
\be
\bm x(t) = \sum_{k=0}^{K-1} \bm p_k s_k(t)
= \sum_{i=-\infty}^{+\infty} \sum_{k=0}^{K-1}
\bm p_k \, b_k(i) \, \psi(t-i \, T_\text s)
\ee
where $\bm p_k \in \Cset^{1 \times N}$ represents the precoding
(beamforming) row-vector associated with the $k$-th data stream.
Leveraging the matrix representation one obtains
\be
\bm x(t) = \sum_{i=-\infty}^{+\infty}
\bm b(i) \, \bm P \, \psi(t-i \, T_\text s)
\label{eq:prec}
\ee
where $\bb(i) \eqdef [b_0(i),b_1(i),\ldots,b_{K-1}(i)] \in \Cset^{1 \times K}$,
and $\bm P \eqdef \left[\bm p^\trasp_0,\bm p^\trasp_1,\ldots,\bm p^\trasp_{K-1}\right]^\trasp \in \Cset^{K \times N}$ denotes the precoding matrix.

The baseband signal radiated from the $L$-th layer of the SIM
and propagating through the physical channel admits the following expression:
\be
\bm z(t) = \bm x(t) \, \bm G = \sum_{n=0}^{N-1} x_n(t) \, \bm g_n.
\label{eq:z}
\ee

Using \eqref{eq:prec} and \eqref{eq:z},
the total power radiated by the SIM can be expressed as
\be
\euscr{P}_{\text{rad}} \eqdef <\Es[\|\bm z(t)\|^2]>
= \|\bm P \bm G\|_\text F^2 \le \euscr{P}_{\text{S}} \|\bm G\|^2
\label{eq:Prad}
\ee
where $\Es[\cdot]$ represents the ensemble average,
$<x(t)> \eqdef \lim_{T\rightarrow+\infty}
\frac{1}{T} \int_{-T/2}^{T/2} x(t)\,dt$ is the time average 
of the continuous-time signal $x(t)$,
$\euscr{P}_{\text{S}} \eqdef \|\bm P\|_\text F^2$ is 
the overall power emitted by the SAT antenna array, 
with $\| \bm A \|_\text F$ denoting the Frobenius norm of the matrix $\bm A$, 
and we have exploited the statistical independence
among the information symbols $b_k(i)$.

At the $k$-th receiver, which is assumed to lie in the far field of the SIM,
the waveform is passed through a matched filter with impulse response
$\psi(-t)$ and uniformly sampled at the symbol rate \(1/T_{\text{s}}\),
assuming perfect timing synchronization.
The resulting baseband signal $y_k(i) \triangleq y_k(i \, T_{\text{s}})$
received by the $k$-th user during the $i$-th symbol interval $[i \, T_{\text{s}},(i+1)\, T_{\text{s}})$, with $k \in \{0,1,\ldots,K-1\}$ and $i \in \Zset$,
is given by
\be
y_k(i) = \bm b(i) \, \bm P \, \bm G \, \bm h_k^\trasp + r_k(i)
\ee
where $\bm h_k \in \Cset^{1 \times N}$ represents
the frequency-flat block-fading channel
from the SIM to the $k$-th user%
\footnote{The considered signal model can be readily adapted to
more complicated modulation schemes (e.g., CP-OFDM)
possibly adopted to cope with the time-dispersive nature of
wireless channels.},
and $r_k(i)$ denotes
(filtered) circularly-symmetric
complex Gaussian noise at the $k$-th user, 
with zero mean and variance $\Es[|r_k(i)|^2] = \sigma_k^2$.

By stacking the received samples of all UEs, we obtain
\be
\mathbf{y}(i) = \mathbf{b}(i)\, \mathbf{P}\, \mathbf{G}\, \mathbf{H} + \mathbf{r}(i)
\label{eq:sig-rx}
\ee
where 
$\mathbf{y}(i) \triangleq [y_0(i),\, y_1(i),\, \ldots,\, y_{K-1}(i)] \in \mathbb{C}^{1 \times K}$,  
$\mathbf{r}(i) \triangleq [r_0(i),\, r_1(i),\, \ldots,\, r_{K-1}(i)] \in \mathbb{C}^{1 \times K}$ denotes the noise vector, and $\mathbf{H} \triangleq [\mathbf{h}_0^\top,\, \mathbf{h}_1^\top,\, \ldots,\, \mathbf{h}_{K-1}^\top] \in \mathbb{C}^{Q^{(L)} \times K}$ is the channel matrix.

The parameters $\mathbf{P}$ and
$\{\boldsymbol{\tau}_\ell\}_{\ell \in \mathcal{L}}$
can be tuned by the designer to 
optimize the communication links.


\section{Model-Based Optimization}
\label{sec:model-opt}

Our aim is to jointly determine the precoding matrix $\mathbf{P}$ 
and the SIM transmission coefficients $\{\tau_{\ell,q}\}$, 
for each layer $\ell$ and meta-atom $q$, 
that minimize the following\emph{mean squared error} (MSE):
\be
\mathrm{MSE}\!\left(\mathbf{P}, \{\boldsymbol{\tau}_\ell\}_{\ell \in \mathcal{L}}\right) 
\triangleq \mathbb{E}\!\left[\|\mathbf{b}(i) - \beta\, \mathbf{y}(i)\|^2\right]
\label{eq:mse}
\ee
subject to the transmit power constraint $\|\mathbf{P}\|^2 = \euscr{P}_{\mathrm{S}}$.
By substituting \eqref{eq:sig-rx} into \eqref{eq:mse}, the optimal solution 
for the precoding matrix can be expressed as \cite{Joham2005}
\be
\mathbf{P}_{\mathrm{MMSE}} = \beta^{-1} \mathbf{H}^\herm \mathbf{G}^\herm
\left(\mathbf{G}\mathbf{H}\mathbf{H}^\herm \mathbf{G}^\herm + \frac{1}{\mathrm{SNR}} \mathbf{I}_N\right)^{-1}
\label{eq:precoding}
\ee
where the signal-to-noise ratio is defined as
\be
\mathrm{SNR} \triangleq \frac{\euscr{P}_{\mathrm{S}}}{\sum_{k=0}^{K-1} \sigma_k^2}
\ee
and the normalization factor $\beta$ is given by
\be
\beta = \sqrt{\frac{\operatorname{tr}\!\left[ \mathbf{H}^\herm \mathbf{G}^\herm
\left(\mathbf{G}\mathbf{H}\mathbf{H}^\herm \mathbf{G}^\herm 
+ \frac{1}{\mathrm{SNR}} \mathbf{I}_N \right)^{-2}
\mathbf{G}\mathbf{H}\right]}{\euscr{P}_{\mathrm{S}}}} \: .
\ee

Substituting \eqref{eq:precoding} into \eqref{eq:mse}, we obtain
\begin{multline}
\mathrm{MSE}\!\left(\mathbf{P}_{\mathrm{MMSE}}, \{\boldsymbol{\tau}_\ell\}_{\ell \in \mathcal{L}}\right) =
K - \operatorname{tr}\!\left[\mathbf{H}^\herm \mathbf{G}^\herm \right. \\ \left.
\left(\mathbf{G}\mathbf{H}\mathbf{H}^\herm \mathbf{G}^\herm + \frac{1}{\mathrm{SNR}} \mathbf{I}_N\right)^{-1}
\mathbf{G}\mathbf{H}\right].
\label{eq:opt-mse}
\end{multline}

By further exploiting the spectral representation of \eqref{eq:opt-mse}, it follows that
\be
\mathrm{MSE}\!\left(\mathbf{P}_{\mathrm{MMSE}}, \{\boldsymbol{\tau}_\ell\}_{\ell \in \mathcal{L}}\right) =
K - \sum_{k=0}^{R_m-1}
\frac{\tilde{s}_k^2}{\tilde{s}_k^2 + 1/\mathrm{SNR}}
\label{eq:mse-spectral}
\ee
where $R_m \triangleq \min\!\big(K, \operatorname{rank}(\mathbf{G}\mathbf{H})\big)$,
and $\tilde{s}_k \in \mathbb{R}$ denotes the $k$-th singular 
value of $\mathbf{G}\mathbf{H}$.  

Assuming that $\mathbf{H}$ is full rank, the objective is to design $\mathbf{G}$
so that $\mathbf{G}\mathbf{H}$ is also full rank, i.e., $R_m = K$, while maximizing the singular values $\{\tilde{s}_k^2\}_{k=0}^{K-1}$ 
under constraint~\eqref{eq:G}.  
This is achieved when the rows of $\mathbf{G}$ are aligned 
with the left singular vectors of $\mathbf{H}$.

Specifically, consider the singular value decomposition (SVD) of  
$\mathbf{H} = \mathbf{U}_\text{H} \mathbf{S}_\text{H} \mathbf{V}_\text{H}^\herm$,  
where $\mathbf{U}_\text{H} \in \mathbb{C}^{Q^{(L)} \times Q^{(L)}}$ and  
$\mathbf{V}_\text{H} \in \mathbb{C}^{K \times K}$ are unitary matrices, and
\be
\mathbf{S}_\text{H} \triangleq
\begin{bmatrix}
s_0 & 0 & \cdots & 0 \\
0 & s_1 & \cdots & 0 \\
\vdots & \vdots & \ddots & \vdots \\
0 & 0 & \cdots & s_{K-1} \\
0 & 0 & \cdots & 0 \\
\vdots & \vdots & \ddots & \vdots \\
0 & 0 & \cdots & 0
\end{bmatrix} \in \mathbb{R}^{Q^{(L)} \times K}.
\ee

The SIM can be designed such that  
$\mathbf{G} = \mathbf{U}_\text{G} \mathbf{D}\, \mathbf{U}_{\text{H},\text{left}}^\herm$,  
where $\mathbf{U}_{\text{H},\text{left}} \in \mathbb{C}^{Q^{(L)} \times N}$ collects the first $N$ columns of $\mathbf{U}_\text{H}$,  
$\mathbf{D} \triangleq \operatorname{diag}(d_0, d_1, \ldots, d_{N-1}) \in \mathbb{R}^{N \times N}$,  
and $\mathbf{U}_\text{G} \in \mathbb{C}^{N \times N}$ is unitary \cite{Horn}.  
For $N \geq K$, the MSE expression in \eqref{eq:mse-spectral} reduces to
\be
\mathrm{MSE}\!\left(\mathbf{P}_{\mathrm{MMSE}}, \{\boldsymbol{\tau}_\ell\}_{\ell \in \mathcal{L}}\right) =
K - \sum_{k=0}^{K-1}
\frac{d_k^2 s_k^2}{d_k^2 s_k^2 + 1/\mathrm{SNR}}.
\label{eq:mse-spectral-opt}
\ee

The optimization problem therefore consists in maximizing the sequence  
$\{d_k\}_{k=0}^{K-1}$ while satisfying the physical constraints imposed by the SIM architecture.  
A common simplification is to impose mild, physically motivated constraints such as  
$\mathbf{G} = \mathbf{U}_{\text{H},\text{left}}^\herm$ \cite{Dar2025},  
which enforce specific structural properties on the rows of $\mathbf{G}$.

It is important to note that the model-based approach outlined above
assumes perfect knowledge at the SAT of both the overall channel $\Hb$ and the SNR.
In practice, these parameters must be estimated,
which not only incurs extra communication overhead
but also introduces additional errors in the reception of information.

In the following sections, we propose a learning-based design
of both the SIM and the precoder, which exploits modern GPU architectures
and the Sionna framework to tackle this optimization problem,
while fully exploiting the resources of the SIM.

\section{Implementation in TensorFlow}
\label{sec:implementation}

In this section, we describe the implementation 
of the SIM response and the precoding matrix within the TensorFlow framework.  
Both have been realized as custom Keras layers \cite{chollet2015keras}, making them readily integrable into Sionna or, more generally, into any TensorFlow-based pipeline.

The following provides the implementation of the \texttt{SIM} class.

\begin{Verbatim}[breaklines=true]
import tensorflow as tf
...
class SIM(tf.keras.layers.Layer):
    def __init__(self, W_list, mask=None, gain_bounds=None, passive_gain=1.0, seed=None, name="SIM", **kwargs):
...
\end{Verbatim}

As shown, the class takes as input the following arguments:  
a list \texttt{W\_list}, which collects the transition matrices 
defined in \eqref{eq:W1} and \eqref{eq:Well};  
a list \texttt{mask}, specifying for each layer whether 
it is \emph{active} or \emph{passive};  
a list \texttt{gain\_bounds}, defining the amplitude bounds for the active layers;  
the floating-point value \texttt{passive\_gain}, representing the gain introduced by the passive layers (i.e., $\alpha_{\text{pc}}$);  
and the integer \texttt{seed}, used to initialize the internal random generator.  

The following provides the definition of the \texttt{TrainablePrecoder} class.

\begin{Verbatim}[breaklines=true]
import tensorflow as tf
...
class TrainablePrecoder(tf.keras.layers.Layer):
    def __init__(self, K, N, Ps, init, name="TrainablePrecoder", **kwargs):
...
\end{Verbatim}

As shown, the class takes as input the following arguments:  
the integers \texttt{K} and \texttt{N}, denoting the number of UEs and the number of antennas, respectively;  
the floating-point value \texttt{P\_s}, representing the total transmit power of the antenna array, i.e., $\euscr{P}_{\mathrm{S}}$;  
and the $K \times N$ complex matrix \texttt{init}, which specifies the initial values of the precoding matrix.  

Both classes have been implemented to be fully differentiable, thereby enabling the exploitation of automatic gradient computation in TensorFlow with GPU acceleration.

\section{Simulation Results}
\label{sec:experiments}

Monte Carlo simulations were carried out
for analyzing the performance of the AI-driven SIM design strategy in comparison with the model-based approach, in terms of \emph{bit-error rate} (BER) versus the energy contrast $E_{\text{b}}/N_0$.

Specifically, we considered a SAT equipped
with an \emph{uniform planar array} (UPA)
composed of $N=4$ antennas, performing 
downlink transmission
of $K=4$ information streams 
to $K=4$ single-antenna UEs.
The UPA antenna spacing is set 
to $d_{\text{S}} = \lambda_0/2$.

The considered SIM consists of $L=8$ equally spaced 
layers with inter-layer spacing $s = \lambda_0/2$, 
of which $L_{\text{ac}} = 2$ are active layers and $L_{\text{pc}} = 6$ 
are passive ones.  
Each layer is formed by $Q = 12 \times 12$ unit cells, arranged with an 
inter-cell spacing of $s = \lambda_0/2$.  
The gain of the active unit cells can be tuned between $-22$ dB 
and $13$ dB, whereas the gain of the passive unit cells is 
fixed to $\alpha_{\text{pc}} = 0.9$.
Quantization effects are not explicitly accounted for,
as the proposed solution is applicable to non-reconfigurable architectures
as well as to reconfigurable architectures employing meta-atoms
that can be controlled in an (almost) continuous manner.

The native \texttt{sionna} classes, \texttt{BinarySource},
\texttt{Mapper}, \texttt{Demapper}, and \texttt{AWGN}
are used for implementing the end-to-end chain.

For each realization, the entries of the channel matrix $\mathbf{H}$ are generated as i.i.d. circularly symmetric complex Gaussian random variables with zero mean and unit variance. In addition, a block of $N_{\text{B}} = 1000$ bits is generated for each user.  
For the data-driven approaches, the matrices $\mathbf{P}$ and $\mathbf{G}$ are learned using a block of $100$ symbols per realization.

\begin{figure}
    \centering
    \includegraphics[width=\linewidth]{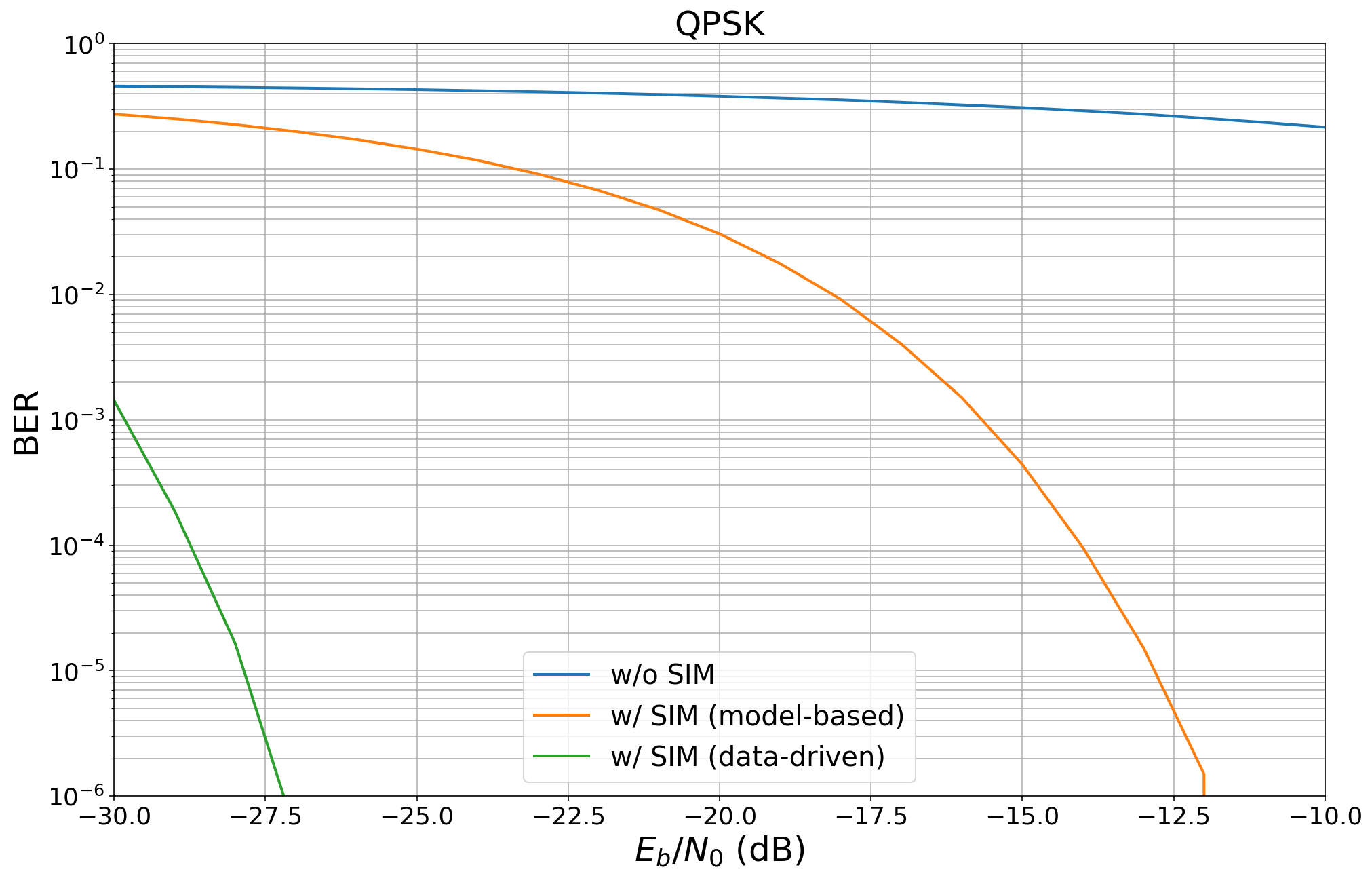}
    \caption{Bit-error rate versus energy contrast for QPSK modulation scheme.}
    \label{fig:qpsk}
\end{figure}

\begin{figure}
    \centering
    \includegraphics[width=\linewidth]{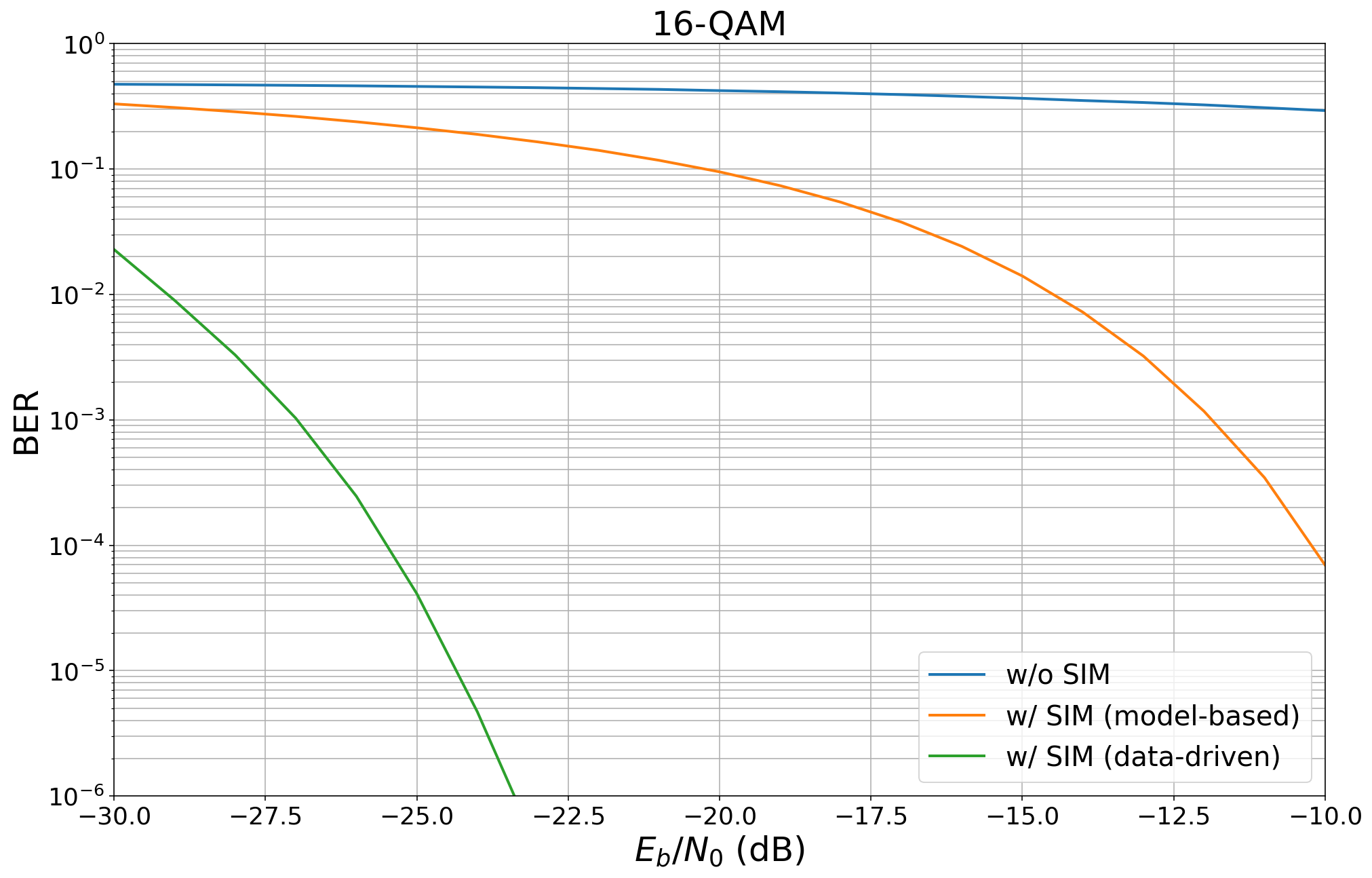}
    \caption{Bit-error rate versus energy contrast for 16-QAM modulation scheme.}
    \label{fig:16qam}
\end{figure}

Figures~\ref{fig:qpsk} and \ref{fig:16qam} report the simulation results averaged over $N_{\text{mc}} = 1000$ realizations, for QPSK and 16-QAM modulation schemes, respectively.  
The results clearly indicate that the data-driven approach (green curves) significantly outperforms both is model-based counterpart (orange) and the baseline system without SIM (blue).  

At first glance, this behavior may seem counterintuitive; however, it can be explained by the inherent limitations of the adopted SIM model.  
In particular, the model-based SIM synthesis is obtained by imposing $\mathbf{G} = \mathbf{U}_{\text{H,left}}^\herm$, which implies $d_k = 1$ for all $k \in \{0,1,\ldots,K-1\}$.  
This assumption does not allow to fully exploit 
the array gain that the SIM could provide, especially when active 
layers are included.  
Conversely, the data-driven approach is able to exploit the 
full set of degrees of control offered by the SIM.  

When only passive layers are considered, one has $d_k < 1$ for all $k \in \{0,1,\ldots,K-1\}$, and in this case the performance of the data-driven approach deteriorates 
compared to the model-based design.  
For brevity, the corresponding simulation results are omitted.

\section{Conclusion and Future Work}
\label{sec:conclusion}

This paper introduced a differentiable and GPU-accelerated model
for the design and evaluation of SIM integrated within NVIDIA’s Sionna library.
By developing a multi-layer SIM model and embedding it into TensorFlow,
we enabled end-to-end optimization of both
the precoder and the metasurface configuration.
Monte Carlo simulations showed that the learning-based design
consistently outperforms its model-based counterpart
as well as a baseline system without SIM,
confirming the benefits of data-driven adaptation
in exploiting the additional DoC
provided by stacked phy\-si\-cal architectures.

The study has been conducted under simplifying assumptions, 
including narrowband block-fading channels and single-carrier modulation. 
These limitations will be addressed in future work by 
leveraging the \texttt{ofdm} sub-module of \texttt{sionna} 
and the OpenNTN framework \cite{OpenNTNPaper}. 
Future extensions include wideband channel modeling, 
robust learning under partial CSI, and online adaptation techniques 
tailored to time-varying NTN scenarios.  

By introducing an open and scalable integration 
of SIM into an AI-native simulator, this work establishes a 
foundation for programmable, software-defined environments in next-generation aerospace and satellite communication systems.

\acknowledgements 
This work was partially supported by the European Union-Next Generation EU
under the Italian National Recovery and Resilience Plan (NRRP),
Mission 4, Component 2, Investment 1.3, CUP E63C22002040007,
partnership on ``Telecommunications of the Future" (PE00000001 - program ``RESTART").

The work of I. Iudice was partially supported by the
Italian National Project ``Maturazione Tecnologie Innovative
Mini e Micro Droni (MATIM)" through
Programma Nazionale di Ricerche Aerospaziali (PRORA) under Grant DM662.

The research activities presented in this paper fall within the field of interest of the IEEE AESS technical panel on Glue Technologies for Space Systems.
%

\bibliographystyle{IEEEtran}
\bibliography{biblio}

@article{Hoy2022,
  author    = {Jakob Hoydis and Sebastian Cammerer and Fay\c{c}al A\"it Aoudia and \emph{et al.}},
  title     = {Sionna: An Open-Source Library for Next-Generation Physical Layer Research},
  journal   = {arXiv preprint},
  year      = {2022},
}

@misc{SionnaSW2025,
  author       = {Hoydis, Jakob and Cammerer, Sebastian and A{\"i}t Aoudia, Fay\c{c}al and Nimier-David, Merlin and Maggi, Lorenzo and Marcus, Guillermo and Vem, Avinash and Keller, Alexander},
  title        = {Sionna},
  howpublished = {\url{https://github.com/NVlabs/sionna}},
  note         = {GPU-accelerated, differentiable 6G research library},
  year         = {2025},
  version      = {1.1.0},
}

@ARTICLE{Pan2020,
  author={Pan, Cunhua and Ren, Hong and Wang, Kezhi and Kolb, Jonas Florentin and Elkashlan, Maged and Chen, Ming and Di Renzo, Marco and Hao, Yang and Wang, Jiangzhou and Swindlehurst, A. Lee and You, Xiaohu and Hanzo, Lajos},
  journal={IEEE Communications Magazine}, 
  title={Reconfigurable Intelligent Surfaces for {6G} Systems: Principles, Applications, and Research Directions}, 
  year={2021},
  volume={59},
  number={6},
  pages={14-20},
  doi={10.1109/MCOM.001.2001076}}

@ARTICLE{Bas2021,
  author={Basharat, Sarah and Hassan, Syed Ali and Pervaiz, Haris and Mahmood, Aamir and Ding, Zhiguo and Gidlund, Mikael},
  journal={IEEE Wireless Communications}, 
  title={Reconfigurable Intelligent Surfaces: Potentials, Applications, 
         and Challenges for {6G} Wireless Networks}, 
  year={2021},
  volume={28},
  number={6},
  pages={184-191},
  doi={10.1109/MWC.011.2100016}}

@Article{Worka2024,
AUTHOR = {Worka, Chika E. and Khan, Faheem A. and Ahmed, Qasim Zeeshan and Sureephong, Pradorn and Alade, Temitope},
TITLE = {Reconfigurable Intelligent Surface ({RIS})-Assisted Non-Terrestrial Network ({NTN})-Based {6G} Communications: A Contemporary Survey},
JOURNAL = {Sensors},
VOLUME = {24},
YEAR = {2024},
NUMBER = {21},
ARTICLE-NUMBER = {6958},
URL = {https://www.mdpi.com/1424-8220/24/21/6958},
PubMedID = {39517855},
ISSN = {1424-8220},
DOI = {10.3390/s24216958}
}

@article{Zha2019,
  author    = {Jun Zhao},
  title     = {A Survey of Intelligent Reflecting Surfaces ({IRSs}): Towards {6G} Wireless Communication Networks},
  journal   = {arXiv preprint},
  year      = {2019},
  eprint    = {1907.04789},
}

@ARTICLE{Liu2020,
  author={Liu, Yuanwei and Liu, Xiao and Mu, Xidong and Hou, Tianwei and Xu, Jiaqi and Di Renzo, Marco and Al-Dhahir, Naofal},
  journal={IEEE Communications Surveys \& Tutorials}, 
  title={Reconfigurable Intelligent Surfaces: Principles and Opportunities}, 
  year={2021},
  volume={23},
  number={3},
  pages={1546-1577},
  doi={10.1109/COMST.2021.3077737}}

@article{Put2025,
  author    = {Prasetyo Putranto and Anis Amazigh Hamza and Sameh Mabrouki and Nasrullah Armi and Iyad Dayoub},
  title     = {Reconfigurable Intelligent Surfaces for {6G} and Beyond: A Comprehensive Survey},
  journal   = {arXiv preprint},
  year      = {2025},
  eprint    = {2506.19526},
}

@ARTICLE{Ume2023,
  author={Umer, Anum and Müürsepp, Ivo and Alam, Muhammad Mahtab and Wymeersch, Henk},
  journal={IEEE Communications Surveys \& Tutorials}, 
  title={Reconfigurable Intelligent Surfaces in {6G} Radio Localization: A Survey of Recent Developments, Opportunities, and Challenges}, 
  year={2025},
  volume={},
  number={},
  pages={1-1},
  doi={10.1109/COMST.2025.3536517}}

@ARTICLE{Wu2021,
  author={Wu, Qingqing and Zhang, Rui},
  journal={IEEE Communications Magazine}, 
  title={Towards Smart and Reconfigurable Environment: Intelligent Reflecting Surface Aided Wireless Network}, 
  year={2020},
  volume={58},
  number={1},
  pages={106-112},
  doi={10.1109/MCOM.001.1900107}}

@misc{NVlab2023,
  author    = {{NVIDIA Labs}},
  title     = {Sionna RT: Differentiable Ray Tracing for Radio Propagation Modeling},
  year      = {2023},
  howpublished = {\url{https://nvlabs.github.io/sionna/}},
}

@ARTICLE{Let2019,
  author={Letaief, Khaled B. and Chen, Wei and Shi, Yuanming and Zhang, Jun and Zhang, Ying-Jun Angela},
  journal={IEEE Communications Magazine}, 
  title={The Roadmap to {6G}: {AI} Empowered Wireless Networks}, 
  year={2019},
  volume={57},
  number={8},
  pages={84-90},
  doi={10.1109/MCOM.2019.1900271}}

@article{Lin.2018,
author = {Xing Lin  and Yair Rivenson  and Nezih T. Yardimci  and Muhammed Veli  and Yi Luo  and Mona Jarrahi  and Aydogan Ozcan },
title = {All-optical machine learning using diffractive deep neural networks},
journal = {Science},
volume = {361},
number = {6406},
pages = {1004-1008},
year = {2018},
doi = {10.1126/science.aat8084},
URL = {https://www.science.org/doi/abs/10.1126/science.aat8084},
eprint = {https://www.science.org/doi/pdf/10.1126/science.aat8084}}

@article{Liu.2022,
  title={A programmable diffractive deep neural network based on a digital-coding metasurface array},
  author={Liu, Che and Ma, Qian and Luo, Zhang Jie and Hong, Qiao Ru and Xiao, Qiang and Zhang, Hao Chi and Miao, Long and Yu, Wen Ming and Cheng, Qiang and Li, Lianlin and others},
  journal={Nature Electronics},
  volume={5},
  number={2},
  pages={113--122},
  year={2022},
  publisher={Nature Publishing Group UK London}
}

@ARTICLE{Hassan.2024,
  author={Hassan, Naveed Ul and An, Jiancheng and Di Renzo, Marco and Debbah, Mérouane and Yuen, Chau},
  journal={IEEE Open Journal of the Communications Society}, 
  title={Efficient Beamforming and Radiation Pattern Control Using Stacked Intelligent Metasurfaces}, 
  year={2024},
  volume={5},
  number={},
  pages={599-611},
  doi={10.1109/OJCOMS.2023.3349155}}

@ARTICLE{Nerini.2024,
  author={Nerini, Matteo and Clerckx, Bruno},
  journal={IEEE Communications Letters}, 
  title={Physically Consistent Modeling of Stacked Intelligent Metasurfaces Implemented With Beyond Diagonal RIS}, 
  year={2024},
  volume={28},
  number={7},
  pages={1693-1697},
  doi={10.1109/LCOMM.2024.3401580}}

@ARTICLE{DiRenzo,
  author={An, Jiancheng and Yuen, Chau and Guan, Yong Liang and Renzo, Marco Di and Debbah, Mérouane and Poor, H. Vincent and Hanzo, Lajos},
  journal={IEEE Journal on Selected Areas in Communications}, 
  title={Two-Dimensional Direction-of-Arrival Estimation Using Stacked Intelligent Metasurfaces}, 
  year={2024},
  volume={42},
  number={10},
  pages={2786-2802},
  doi={10.1109/JSAC.2024.3414613}}

@ARTICLE{Hanzo,
  author={An, Jiancheng and Xu, Chao and Ng, Derrick Wing Kwan and Alexandropoulos, George C. and Huang, Chongwen and Yuen, Chau and Hanzo, Lajos},
  journal={IEEE Journal on Selected Areas in Communications}, 
  title={Stacked Intelligent Metasurfaces for Efficient Holographic {MIMO} Communications in 6G}, 
  year={2023},
  volume={41},
  number={8},
  pages={2380-2396},
  doi={10.1109/JSAC.2023.3288261}}

@INPROCEEDINGS{DiRenzo-ICC,
  author={An, Jiancheng and Di Renzo, Marco and Debbah, Mérouane and Yuen, Chau},
  booktitle={ICC 2023 - IEEE International Conference on Communications}, 
  title={Stacked Intelligent Metasurfaces for Multiuser Beamforming in the Wave Domain}, 
  year={2023},
  volume={},
  number={},
  pages={2834-2839},
  doi={10.1109/ICC45041.2023.10279173}}

@INPROCEEDINGS{Liu.2024,
  author={Liu, Hao and An, Jiancheng and Ng, Derrick Wing Kwan and Alexandropoulos, George C. and Gan, Lu},
  booktitle={ICC 2024 - IEEE International Conference on Communications}, 
  title={{DRL}-Based Orchestration of Multi-User {MISO} Systems with Stacked Intelligent Metasurfaces}, 
  year={2024},
  volume={},
  number={},
  pages={4991-4996},
  doi={10.1109/ICC51166.2024.10622385}}

@ARTICLE{Yao.2024,
  author={Yao, Xianghao and An, Jiancheng and Gan, Lu and Di Renzo, Marco and Yuen, Chau},
  journal={IEEE Wireless Communications Letters}, 
  title={Channel Estimation for Stacked Intelligent Metasurface-Assisted Wireless Networks}, 
  year={2024},
  volume={13},
  number={5},
  pages={1349-1353},
  doi={10.1109/LWC.2024.3369874}}

@ARTICLE{Lin.2024,
  author={Lin, Shining and An, Jiancheng and Gan, Lu and Debbah, Mérouane and Yuen, Chau},
  journal={IEEE Wireless Communications Letters}, 
  title={Stacked Intelligent Metasurface Enabled {LEO} Satellite Communications Relying on Statistical {CSI}}, 
  year={2024},
  volume={13},
  number={5},
  pages={1295-1299},
  doi={10.1109/LWC.2024.3368238}}

@book{Goodman,
  title={Introduction to Fourier optics},
  author={Goodman, Joseph W},
  year={2005},
  publisher={Roberts and Company publishers}
}

@book{Horn,
  title={Matrix Analysis},
  author={Horn, R. A. \& Johnson, C. R.},
  year={1985},
  publisher={Cambridge University Press, Cambridge.}
}

@ARTICLE{Dar2025,
  author={Darsena, Donatella and Verde, Francesco and Iudice, Ivan and Galdi, Vincenzo},
  journal={IEEE Open Journal of the Communications Society}, 
  title={Design of Stacked Intelligent Metasurfaces With Reconfigurable Amplitude and Phase for Multiuser Downlink Beamforming}, 
  year={2025},
  volume={6},
  number={},
  pages={531-550},
  doi={10.1109/OJCOMS.2025.3526126}}

@Article{Joham2005,
  author    = {M. Joham and W. Utschick and J.A. Nossek},
  journal   = {IEEE Transactions on Signal Processing},
  title     = {Linear transmit processing in {MIMO} communications systems},
  year      = {2005},
  issn      = {1941-0476},
  month     = aug,
  pages     = {2700--2712},
  volume    = {53},
  doi       = {10.1109/TSP.2005.850331},
  issue     = {8},
  publisher = {IEEE},
}

@misc{chollet2015keras,
  title={Keras},
  author={Chollet, Fran\c{c}ois and others},
  year={2015},
  howpublished={\url{https://keras.io}},
}

@inproceedings{OpenNTNPaper,
author = {T. D\"{u}e and M. Vakilifard and C. Bockelmann and D. W\"{u}bben and A. Dekorsy},
year = {2025},
month = {Feb},
title = {An Open Source Channel Emulator for Non-Terrestrial Networks},
URL = {https://www.ant.uni-bremen.de/sixcms/media.php/102/15080/An%20Open%20Source%20Channel%20Emulator%20for%20Non-Terrestrial%20Networks.pdf},
address={Sitges, Spain},
booktitle={Advanced Satellite Multimedia Systems Conference/Signal Processing for Space Communications Workshop (ASMS/SPSC 2025)}
}

\begin{biographywithpic}
{Ivan Iudice}{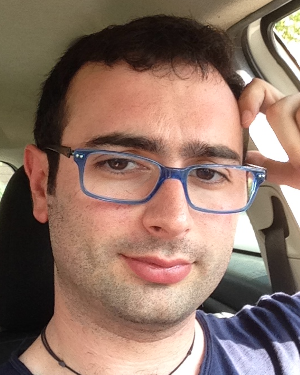}
was born in Livorno, Italy, in 1986.
He received the B.S. and M.S. degrees in telecommunications engineering
in 2008 and 2010, respectively, and the Ph.D. degree in
information technology and electrical engineering in 2017,
all from University of Napoli Federico II, Italy.
Since 2011, he has been with the Italian Aerospace Research Centre (CIRA),
Capua, Italy. He first served as part of the
Electronics and Communications Laboratory and
he is currently part of the Security Unit.
His research activities mainly lie in the area of
signal and array processing for communications,
with current interests focused on physical-layer security,
space-time techniques for cooperative communications systems and
reconfigurable metasurfaces.
He is involved in several international projects.
He serves as reviewer for several international journals and as
TPC member for several international conferences.
He is author of several papers on refereed journals and international conferences.
He has been serving as an Associate Editor for IEEE SIGNAL PROCESSING LETTERS
since 2025.
\end{biographywithpic} 

\begin{biographywithpic}
{\noindent Giacinto Gelli}{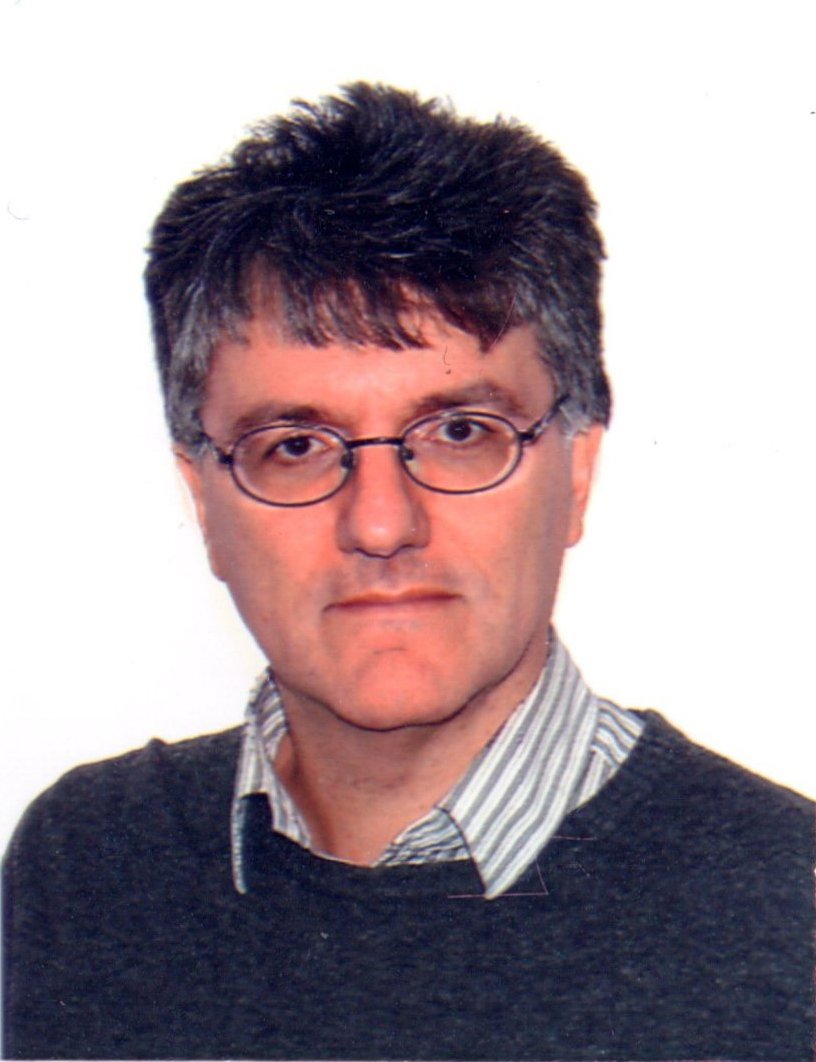} was born in Napoli, Italy, in 1964.
He received the Dr. Eng. degree {\em summa cum laude}
in electronic engineering in 1990,
and the Ph.D. degree in computer science and
electronic engineering in 1994,
both from the University of Napoli
Federico II, Italy.
From 1994 to 1998, he was an Assistant Professor with the
Department of Information Engineering, Second University of
Napoli, Italy. Since 1998 he has been with the Department
of Electrical Engineering and Information Technology,
University of Napoli Federico II, Italy, first as an Associate Professor,
and since November 2006 as a Full Professor of Telecommunications.
He also held teaching positions at the University
Parthenope of Napoli, Italy and 
at the Accademia Aeronautica of Pozzuoli, Italy.
His research interests are in the broad area of
signal and array processing for communications,
with current emphasis on multicarrier modulation systems,
space-time techniques for cooperative and cognitive
communications systems, and backscatter communications.
\end{biographywithpic}

\begin{biographywithpic}{Donatella Darsena}{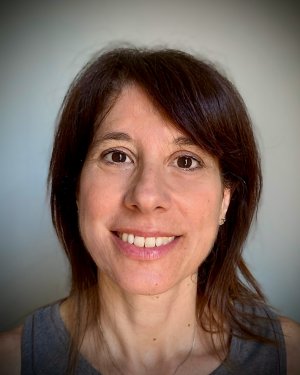}
received the Dr. Eng. degree (summa cum laude) in telecommunications engineering
and the Ph.D. degree in electronic and telecommunications engineering
from the University of Napoli Federico II, Italy, in 2001 and 2005, respectively,
where she is currently an Associate Professor
with the Department of Electrical Engineering and Information Technology.
From 2001 to 2002, she worked as an Embedded System Designer
with the Telecommunications, Peripherals and Automotive Group,
STMicroelectronics, Milano, Italy.
In 2005, she joined the Department of Engineering,
Parthenope University of Napoli, Italy, and worked first
as an Assistant Professor and then as an Associate Professor from 2005 to 2022.
Her research interests are in the broad area of
signal processing for communications,
with current emphasis on reflected-power communications,
orthogonal and nonorthogonal multiple access techniques,
wireless system optimization, and physical-layer security.
She has served as a Senior Editor for IEEE ACCESS since 2024,
as Executive Editor for IEEE COMMUNICATIONS LETTERS since 2023,
and as Associate Editor for IEEE SIGNAL PROCESSING LETTERS since 2020.
She was an Associate Editor for IEEE ACCESS from 2018 to 2023,
and for IEEE COMMUNICATIONS LETTERS from 2016 to 2019,
and a Senior Area Editor for IEEE COMMUNICATIONS LETTERS from 2020 to 2023
\end{biographywithpic}

\end{document}